# Room-temperature ferromagnetism in dielectric GaN(Gd).

V.I. Litvinov[*],

Sierra Nevada Corporation, 15245 Alton Pkwy, Suite 100, Irvine, CA 92618, USA

V.K. Dugaev

Department of Physics, Rzeszow University of Technology

ul. Wincentego Pola 2, 35-959 Rzeszów, Poland;

Department of Physics and CFIF, Instituto Superior Tecnico,

TU Lisbon, Av. Rovisco Pais, 1049-001 Lisbon, Portugal

Abstract.

We present an explanation of recently observed giant magnetic moment and room-temperature ferromagnetism in the dielectric GaN doped with Gd. Our approach uses the polarization mechanism of exchange interaction, which occurs if the d-level of Gd appears in the bandgap close to the valence band edge. Calculated ferromagnetic critical temperature and the value of the magnetic moment well correspond to experimental findings.

---

[*] E-mail: vlitvinov@earthlink.net

Search for the room-temperature ferromagnetism (RTFM) in doped wide bangap AlGaN structures, is an important part of semiconductor spintronics.[1,2,3] It targets the methods of spin injection and electrically-driven spin manipulation in GaN-based such as lasers, light-emitting diodes, and high-power field-effect transistors. Doping with optically active rare-earth elements may introduce additional functionality to the material system since combined magnetic and optical devices may be possible to fabricate on a single chip.

RTFM in wurtzite GaN(Gd) occurs in almost insulating p-type samples.[4,5,6,7] It means that there are no particles at the Fermi level mediating the interspin interaction, in other words, Ruderman-Kittel-Kasuya-Yoshida (RKKY) interaction cannot be at the origin of the ferromagnetism. Moreover, the ferromagnetism was not observed in a low-resistive n-type zinc-blende GaN(Gd).[8] So, the increase in a magnetic moment and a critical temperature associated with the increase in density of Ga vacancies, remains to be explained in terms other than increasing free carrier density.

Doping with non-magnetic atoms strongly affects the ferromagnetism in the wide bandgap nitrides[9,10] If the Fermi level crosses the d-level of a magnetic component, it is important to account for the mixed valence and variable magnetic moment of the magnetic impurity. The mixed valence was taken into account in degenerate semiconductors in Ref.[11].

In this paper we study the ferromagnetic transition in GaN(Gd) assuming that virtual excitations from the Gd d-band to the valence band mediates the Gd-Gd exchange interaction[12,13,14]. The mechanism accounts for following features associated with RTFM in rare-earth doped GaN: the mechanism does not imply free carriers to exist; on the contrary to the double exchange mechanism, it does not require the Fermi level to lie inside the impurity band; it includes the total angular moment comprising f- and d-electron contributions and the variable Gd-Gd coupling as the Fermi level changes its position in the bandgap relatively to the Gd d-band; it allows calculating the giant effective magnetic moment of single Gd impurity embedded in the host.

Gd has both partially filled *4f* and *5d* orbitals. When the density of Ga vacancies increases, it pushes the Fermi level in energy closer to VBM, so two valence states $Gd^{3+}(4f^7)$ and $Gd^{2+}(4f^75d^1)$ exist in the bandgap if the Fermi level crosses the $t_2$ multiplet.

Total moment of the $Gd^{3+}$ is $J_0 = 7/2$ ($L = 0$, $S = 7/2$). In a bivalent state $Gd^{2+}$ the $t_2$ multiplet carries an angular moment of $J_1 = 3/2$ ($L = 1$, $S = 1/2$).

Magnetic part of the impurity Hamiltonian that accounts for the localized $t_2$ level in the bandgap $E_t$ and for valence transitions $Gd^{3+} \leftrightarrow Gd^{2+}$, is written below:

$$H = h_0 J_0^z + (E_t - \mu + h_1 J_1^z)\hat{n}, \tag{1}$$

where $\hat{n} = a^+ a$, $a^+$ is the electron creation operator in the $t_2$-orbital with eigenvalues $n = 0,1$, $\mu$ is the chemical potential, $h_{0,1} = \mu_B g_{0,1} B_{ex}$, $\mu_B$ is the Bohr's magneton, $g_n$ ($g_0 = 1/3$, $g_1 = 1$) and $\hat{J}_n^z$ are the g-factors and the components of a $t_2$-orbital total angular momentum in the direction of an external field $B_{ex}$, respectively. We assume that the Fermi level position $\mu$ is determined by Ga-vacancies and other defects and (or) impurities, so we keep it as a parameter that describes co-doping effects.

Hamiltonian Eq.(1) describes the variable local angular moment that comprises the contributions from either both f-orbital and $t_2$ subshell of the d-orbital ($Gd^{2+}$) or f-orbital only ($Gd^{3+}$).

From the Hamiltonian Eq.(1) one can find the $t_2$-orbital filling factor and magnetization:

$$f = <n> = \left[1 + \frac{sh[\beta h_1/2]\exp[-\beta\varepsilon]}{sh[\beta h_1(J_1+1/2)]}\right]^{-1}, \quad M = f\mu_B g_1 B_{J_1}(h_1) + \mu_B g_0 B_{J_0}(h_0), \tag{2}$$

where $B_J(h) = (J+1/2)cth[(J+1/2)\beta h] - \frac{1}{2}cth[\beta h/2]$, $\varepsilon = \mu - E_t$, $\beta = 1/T$ is the inverse temperature in energy units.

When the magnetic field tends to zero, the filling factor Eq.(2) becomes the Fermi function of the $2J_1 + 1$-fold degenerate level. Magnetic moment Eq.(2) accounts for both f- and d-levels and depends on the filling factor of the $t_2$-orbital that reflects the fact that the co-doping changes the total magnetic moment of Gd.

First-principle calculations place the f-electron density within the valence band very close to VBM.[15,16] There is a strong intra-atomic coupling between d- and f-states making the total Gd magnetic moment being comprised of contributions from both f- and d-shells. The key assumption of the present work is that $t_2$-multiplet of the d-level lies in the bandgap being separated from VBM by a small energy gap $\Delta$ ($\Delta \approx 0.2\,eV$ is taken for numerical example below). If the Fermi level is in the bandgap between VBM and d-level (insulator at T=0), the Gd-Gd exchange interaction can be originates from valence band- d-band virtual transitions across the energy gap $\Delta$:

$$H_{int} = -V(R)\vec{J}(0)\vec{J}(R),$$
$$V(R) = V_0(1-f)\left(\frac{R_0}{R+R_0}\right)^{5/2} \exp[-R/R_0] - kV_0\left(\frac{R_0}{R+R_0}\right)^{10}; \quad (3)$$

where $V_0$ is the interaction energy parameter, $R_0 = \hbar(2\,m_v\,\Delta)^{-1/2}$, $m_v \approx 0.4 m_0$.

First term in Eq.(3) is the interaction mediated by excitations across the energy gap. Factor $1-f$ accounts for the fact that the interaction either exists, if the $t_2$-multiplet is completely or partially empty, or tends to zero as the $t_2$-level becomes fully occupied. Second term in Eq.(4) describes a superexchange contribution that becomes a leading term at short distances $R \ll R_0$. The range function Eq.(3) is shown in Fig.1.

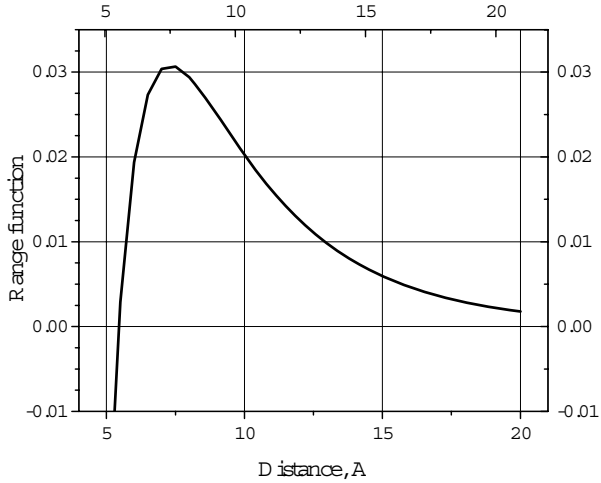

Fig.1. Range function of the Gd-Gd exchange interaction, $\varepsilon = -0.1\,eV$, $T = 300\,K$.

Coefficient $k = 35$ in Eq.(3) is chosen to provide an antiferromagnetic alignment at small distances of order of the lattice constant.

Typical samples contain 1-10% Ga sites occupied by Gd ($x = 0.01 - 0.1$). At $x = 0.05$, the average Gd-Gd distance $R_{av} = \left( \dfrac{3^{4/3} a^2 c}{16 \pi x} \right)^{1/3}$ is equal to 5 Å ($a,c$ are lattice constants in GaN), whereas $R_0$ is about 7 Å. The mean-field approximation holds under the condition $R_0 \gg R_{av}$. As it does not apply to this case, we will use the percolation approach in order to find the critical temperature.

For the pair of Gd atoms on a distance R, coupled by ferromagnetic indirect exchange interaction $V(R)$, the magnetic moment Eq.(3) modifies $h_{0,1} -> h_{0,1} + \dfrac{V(R)M}{\mu_B g_{0,1}}$ and then determines the maximum distance R on which Gd total moments remain parallel:

$$T = \frac{V(R)}{3} \frac{J_0(J_0+1)\exp(-\beta\varepsilon) + (1+2J_1)\left[J_0(1+J_0) + J_1(1+J_1)\right]}{1 + 2J_1 + \exp(-\beta\varepsilon)} \quad (4)$$

The solution to Eq.(4) presents the linear size of the ferromagnetic cluster $R(T)$. Size of the cluster grows with decreasing temperature, and the percolation occurs at the ferromagnetic critical temperature when $R(T)$ reaches the threshold value $R(T_c) = \sqrt[3]{2.4}\, R_{av}$.

Calculated critical temperature is shown in Fig.2. We assume $V_0 = 0.3\, eV$ and keep the Fermi level position as a parameter, $\varepsilon = \mu - E_t$.

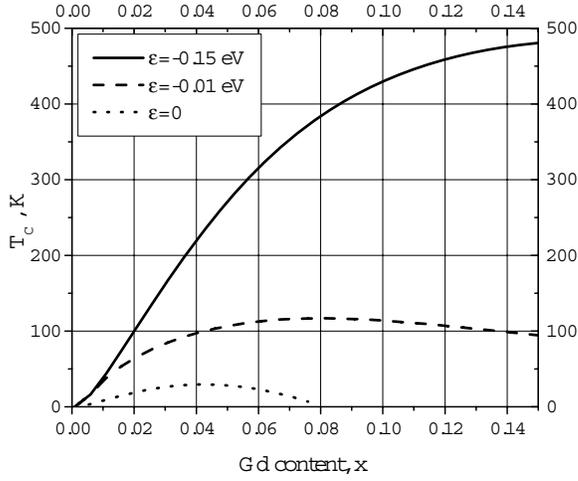

Fig.2. Ferromagnetic critical temperature vs Gd-content.

Fig.2 illustrates the role the co-doping (parameter $\varepsilon$) plays in ferromagnetism. As $\varepsilon$ increases (the n-type co-doping fills out the d-level) the critical temperature drops down to zero. It should be noted that the analytical dependence shown in Fig.2 are qualitatively similar to the first-principle results for transition metal-doped nitrides.[17]

Single Gd impurity polarizes the valence band electrons thus creating large effective magnetic moment $J_{tot}$. The polarization takes place in the range of the interaction $V(R)$ and the effective moment Gd+electrons can be written as

$$J_{tot} = J_{0,1} + J_{ind}, \quad J_{ind} = \frac{J_{0,1}}{2g_{0,1}\mu_B^2\mu_0} \int_{R_0}^{\infty} V(R) d^3R, \qquad (5)$$

where $\mu_0$ is the vacuum magnetic permeability. The moment, induced by the valence electrons, is shown in Fig.3.

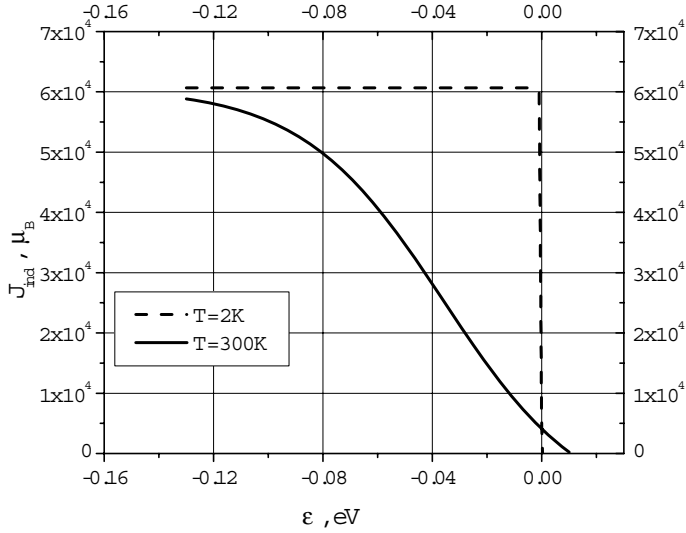

Fig.3. Induced magnetic moment vs Fermi level position.

As shown in Fig.3, the induced moment depends on temperature and Fermi level position, spans the range from zero to $6*10^4$, and tends to zero when $t_2$-band is filled up. If the Fermi level lies slightly below the $t_2$ multiplet ($\varepsilon < 0$), the moment varies between $10^3$ at T=300K to $6*10^4$ at T=2K, that is close to the experimental findings reported in Ref.[6].

The Fermi level location in the energy gap takes place in highly resistive samples which nevertheless reveal ferromagnetic behavior. The results shown in Fig.2, might explain why ferromagnetism was not observed in the n-type zinc-blende GaN. From our perspective, the main reason is a n-type doping that pushes the Fermi level toward the conduction band, makes the $t_2$- multiplet fully occupied, thus making the ferromagnetic Gd-Gd exchange interaction Eq.(3) ineffective. This means that, the controversy as for the very existence of ferromagnetism in GaN(Gd) (paramagnetic behavior has been observed in Refs. [18,19]) should be considered taking into account the accompanying co-doping, more specifically, the position of the Fermi level relative to the $t_2$ multiplet.

Experimentally, the movement of the Fermi level can be achieved in GaN(Gd) epitaxial layers by doping with Si donors in a controllable manner or manipulating the gate voltage in a structure where GaN(Gd) forms a channel of the field-effect transistor.

In conclusion, we present a model of ferromagnetic transition that accounts for variable both the magnetic moment of an individual Gd impurity and Gd-Gd exchange interaction as the Fermi level changes its position in the bandgap as a result of co-doping. Our approach does not exclude the RKKY-mechanism of ferromagnetism that occurs in highly n(p)-doped samples where the Fermi level appears deeply in the conduction (valence) band.

This work is supported by the FCT Grant PTDC/FIS/70843/2006 (Portugal), and by the Polish MNiSW as research projects in years 2006-2010.